\documentclass[sigconf]{acmart}
\AtBeginDocument{%
  \providecommand\BibTeX{{%
    \normalfont B\kern-0.5em{\scshape i\kern-0.25em b}\kern-0.8em\TeX}}}

\copyrightyear{2024} 
\acmYear{2024} 
\setcopyright{acmlicensed}
\acmConference[SIGIR '24]{Proceedings of the 47th International ACM SIGIR Conference on Research and Development in Information Retrieval}{July 14--18, 2024}{Washington, DC, USA}
\acmBooktitle{Proceedings of the 47th International ACM SIGIR Conference on Research and Development in Information Retrieval (SIGIR '24), July 14--18, 2024, Washington, DC, USA}
\acmDOI{10.1145/3626772.3657936}
\acmISBN{979-8-4007-0431-4/24/07}

\usepackage[utf8]{inputenc} 
\usepackage[T1]{fontenc}    
\usepackage{hyperref}       
\usepackage{url}            
\usepackage{booktabs}       
\usepackage{amsfonts}       
\usepackage{nicefrac}       
\usepackage{microtype}      
\usepackage{xcolor}         
\usepackage{graphicx}
\usepackage{amsthm}
\usepackage{amsmath}
\usepackage{subfigure}
\usepackage{cleveref}
\usepackage{caption}
\usepackage{threeparttable}
\usepackage{comment}
\usepackage{multirow}
\usepackage{booktabs}
\usepackage{enumitem}
\usepackage{wrapfig}

\usepackage{adjustbox}
\usepackage{bbding}
\usepackage[normalem]{ulem}
\useunder{\uline}{\ul}{}

\begin{document}

\title{ReCODE: Modeling Repeat Consumption with Neural ODE}

\author{Sunhao Dai}
\author{Changle Qu}
\affiliation{
  \institution{\mbox{Gaoling School of Artificial Intelligence}\\Renmin University of China}
\city{Beijing}
  \country{China}
  }
\email{{sunhaodai, changlequ}@ruc.edu.cn}


\author{Sirui Chen}
\affiliation{
  \institution{University of Illinois at Urbana-Champaign}
\city{Champaign}
  \country{USA}
  }
\email{chensr16@gmail.com}

\author{Xiao Zhang}
\author{Jun Xu}
\authornote{Jun Xu is the corresponding author. Work partially done at Engineering Research Center of Next-Generation Intelligent Search and Recommendation, Ministry of Education.}
\affiliation{%
  \institution{\mbox{Gaoling School of Artificial Intelligence}\\Renmin University of China}
  \city{Beijing}
  \country{China}
  }
\email{{zhangx89,junxu}@ruc.edu.cn}


\renewcommand{\authors}{Sunhao Dai, Changle Qu, Sirui Chen, Xiao Zhang and Jun Xu}
\renewcommand{\shortauthors}{Sunhao Dai, Changle Qu, Sirui Chen, Xiao Zhang and Jun Xu}
\renewcommand{\shorttitle}{ReCODE: Modeling Repeat Consumption with Neural ODE}

\begin{abstract}

In real-world recommender systems, such as in the music domain, repeat consumption is a common phenomenon where users frequently listen to a small set of preferred songs or artists repeatedly. The key point of modeling repeat consumption is capturing the temporal patterns between a user's repeated consumption of the items. Existing studies often rely on heuristic assumptions, such as assuming an exponential distribution for the temporal gaps. However, due to the high complexity of real-world recommender systems, these pre-defined distributions may fail to capture the intricate dynamic user consumption patterns, leading to sub-optimal performance. Drawing inspiration from the flexibility of neural ordinary differential equations (ODE) in capturing the dynamics of complex systems, we propose ReCODE, a novel model-agnostic framework that utilizes neural ODE to model repeat consumption. ReCODE comprises two essential components: a user's static preference prediction module and the modeling of user dynamic repeat intention. By considering both immediate choices and historical consumption patterns, ReCODE offers comprehensive modeling of user preferences in the target context. Moreover, ReCODE seamlessly integrates with various existing recommendation models, including collaborative-based and sequential-based models, making it easily applicable in different scenarios. Experimental results on two real-world datasets consistently demonstrate that ReCODE significantly improves the performance of base models and outperforms other baseline methods.

\end{abstract}




\maketitle

\section{Introduction}

\begin{figure}[t]
  \subfigure[Item A (id: 141574)]
      {
    \includegraphics[width=0.45\columnwidth]{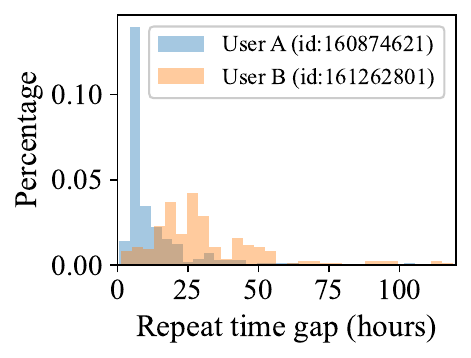}
    \label{fig:intro_stat_user}
  }
  \subfigure[User A (id: 160874621)]{
    \includegraphics[width=0.45\columnwidth]{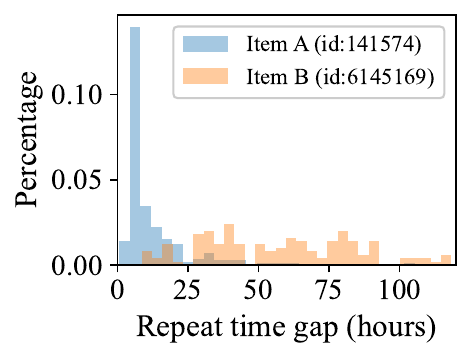}
    \label{fig:intro_stat_item}
  }
  \caption{Repeat consumption time gap distributions of different user-item pairs on MMTD dataset.}
    \label{fig:intro_stat}
\end{figure}

Nowadays, recommender systems have become integral to online service platforms, helping users to cope with information overload~\cite{resnick1997recommender, zhang2019deep, zhang2022counteracting}. Though traditional recommender systems focus on suggesting novel items that users have not yet consumed, repeat consumption has emerged as a prevalent phenomenon in various domains~\cite{benson2016modeling, anderson2014dynamics, teevan2006history, bhagat2018buy, ren2019repeatnet}. For instance, in music recommendation, users often find pleasure in listening to the same songs at different times and in different contexts~\cite{schedl2018current, benson2016modeling, reiter2021predicting, dai2024model}. It is crucial for recommender systems to consider the repeatedly consumed items. 

The key to modeling repeat consumption lies in capturing the pattern of time gaps between a user's consumption of the same items~\cite{benson2016modeling, anderson2014dynamics}. Some approaches, such as ReCANet~\cite{ariannezhad2022recanet} and NovelNet~\cite{li2022modeling}, directly concatenate the temporal gaps with user-item embeddings as input features. On the other hand, other studies often rely on assumptions about the temporal dynamics of repeat consumption, such as empirically assuming an exponential drop-off in the time gap between re-consumptions~\cite{dorogovtsev2000evolution, anderson2014dynamics, reiter2021predicting}. To explicitly capture the diverse temporal dynamics of repeat consumption, Wang et al.~\cite{wang2019modeling} further introduce a mixture distribution approach SLRC, which models short-term effects using an exponential distribution and life-time effects using a Gaussian distribution.

These approaches, although have been commonly adopted and shown their effectiveness, still do not adequately capture the complex and diverse temporal dynamics observed in real-world recommendation scenarios. As illustrated in ~\autoref{fig:intro_stat}, it is evident that different users exhibit different patterns for the same item, and even the same user can have widely different patterns for different items. 
Moreover, this empirical analysis shows that it is difficult to well capture the complex temporal dynamics using a single specific distribution, such as the exponential distribution. 
As a result, the heuristic assumptions made by existing studies~\cite{dorogovtsev2000evolution, anderson2014dynamics, reiter2021predicting, wang2019modeling} do not hold in real-world recommender systems. This limitation hampers the ability to capture fine-grained dynamic properties.

Inspired by the success of neural ordinary differential equations (ODE)~\cite{chen2018neural, cui2023robustness}, this paper proposes to capture the temporal dynamics of repeat consumption patterns in recommendation with ODE, called \textbf{ReCODE}. Compared with existing methods, ReCODE models the consumption patterns with learnable neural networks, without relying on any predefined distributions. In this way, it effectively addresses the aforementioned limitation from the heuristic assumption on the distributions. 
Specifically, ReCODE is designed as a model-agnostic framework and can be implemented based on any existing recommendation models (i.e., the collaborative-based or sequential-based models) as its backbone. It consists of two modules: a static recommendation module that captures the basic user preferences for different items, and a dynamic repeat-aware module that models the temporal patterns of repeat consumption for each user-item pair. By considering both the user's static preference and dynamic repeat intention, ReCODE provides a comprehensive modeling approach for user dynamic preferences. 

The main contributions of this paper are summarized as follows:

$\bullet$ We propose a novel approach called ReCODE for modeling repeat consumption in recommendation. ReCODE can effectively capture the temporal dynamics of user consumption of items without relying on any distribution assumption.

$\bullet$ We implement ReCODE in a model-agnostic manner, which can be easily plugged into various recommendation models, including collaborative-based and sequential-based models. 

$\bullet$ Experimental results and empirical analysis on two real-world datasets demonstrate that ReCODE can consistently enhance various existing recommendation models and is superior to previous baseline models for repeat consumption.

\section{Preliminaries}

\subsection{Problem Formulation}
The user and item sets are denoted as $\mathcal{U}$ and $\mathcal{I}$, respectively.
Each user $u \in \mathcal{U}$ has a chronologically ordered historical consumption sequence $\mathcal{S}_u = \left\{ \left(i_1, t_1\right), \ldots, \left(i_{N_{u}}, t_{N_{u}}\right) \right\}$, where $N_u$ is the number of historical interactions of user $u$ and $i_n \in \mathcal{I}$ is the $n$-th item that the user $u$ has interacted with at time $t_n \in \mathbb{R}^+ $. 
Given the history $\mathcal{S}_u$ of user $u$ and target time $t_{N_u + 1}$, the goal is to predict the next item $i_{N_u + 1} \in \mathcal{I}$ that the user is likely to interact with at $t_{N_u + 1}$.

\subsection{Neural Ordinary Differential Equations}\label{sec:NODE}

Neural ODE \cite{chen2018neural} is a continuous-time model that leverages ordinary differential equations to describe the evolution of a hidden state over time. This framework provides a powerful and flexible approach for capturing the dynamics of complex systems. The key idea behind neural ODE is to represent the hidden state $\boldsymbol{h}(t)$ as a solution to an initial-value problem of an ODE:
$
\frac{d \boldsymbol{h}(t)}{d t}=f_\theta(\boldsymbol{h}(t), t),
$
where the ODE function $f_\theta$ parameterized by $\theta$ is a neural network to approximate the time-derivative of the hidden state. Given the initial state $\boldsymbol{h}({t_0})$, the hidden state $\boldsymbol{h}(t_i)$ can be computed  at any desired time points using a numerical ODE solver:
\begin{equation}
\label{eq:ode_solver}
    \boldsymbol{h}(t_i) = \boldsymbol{h}\left(t_0\right)+\int_{t_0}^{t_i}\boldsymbol{f}_\theta(\boldsymbol{h}(t), t) dt = \textbf{ODESolve}\big(f_{\theta}, \boldsymbol{h}({t_0}), t_i\big),
\end{equation}
where `$\textbf{ODESolve}$' is the ODE solver which discretizes the target time $t_i$ and approximates the integral by performing multiple steps of additions to compute the hidden state $\boldsymbol{h}(t_i)$. Commonly used ODE solvers include fix-step methods~\cite{runge1895numerische} like explicit Euler and fourth-order Runge-Kutta (RK4) as well as adaptive step methods such as Dormand–Prince~\cite{dormand1980family}.

\begin{figure}[t]  
    \centering    
    \includegraphics[width=1\columnwidth]{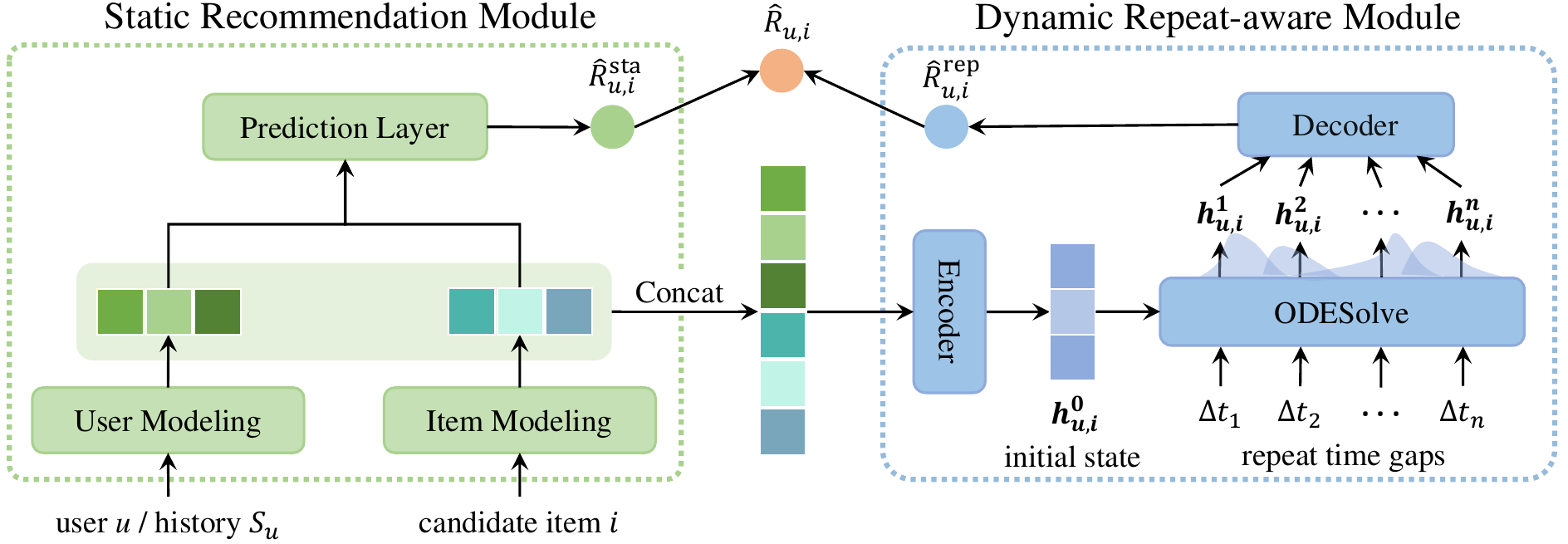}
    \caption{Architecture of our proposed ReCODE framework.}
    \label{fig:model_framework} 
\end{figure}

\section{Our Approach}
In the section, we propose a model-agnostic framework for modeling \textbf{Re}peat \textbf{C}onsumption with neural \textbf{ODE} (denoted as ReCODE). 

\subsection{Framework Overview}
As illustrated in \autoref{fig:model_framework}, our proposed ReCODE framework consists of two main modules: a static recommendation module and a dynamic repeat-aware module. The static recommendation module first takes the user $u$ (or user history interactions $\mathcal{S}_u$) and item $i$ as input to get the user and item representations for predicting the base static user's preference $\hat{R}_{u,i}^\text{sta}$. The combination of the user-item representation is then fed into the repeat-aware module to model the dynamic pattern of the history repeat consumption time intervals and output the repeat intention $\hat{R}_{u,i}^\text{rep}$. Finally, we generate the final prediction score through integrating both the base static user preference and the dynamic repeat intention as
\begin{equation} \label{eq: final_prediction}
    \hat{R}_{u,i} = \hat{R}_{u,i}^\text{sta} + \hat{R}_{u,i}^\text{rep}.
\end{equation}

In the following sections, we will present the details of the two modules to obtain $\hat{R}_{u,i}^\text{sta}$ and $\hat{R}_{u,i}^\text{rep}$, respectively.
\subsection{Static Recommendation Module}

As shown in the left part of \autoref{fig:model_framework}, many recommendation models share a similar two-tower structure~\cite{xu2018deep, chen2018narra, he2017neural, kang2018self}, which we can use as the static recommendation module to model the user’s static preference. Specifically, the static recommendation module predicts the static user preference based on the learned user representation $\boldsymbol{e}_u$ and item representations $\boldsymbol{e}_i$:
\begin{equation}
    \hat{R}_{u,i}^\text{sta} = f_\text{pred} \left(\boldsymbol{e}_u, \boldsymbol{e}_i \right) .
\end{equation}
where $f_\text{pred}$ is the prediction layer and commonly defined as dot product~\cite{koren2009matrix, kang2018self, he2020lightgcn} or fully-connected layers~\cite{he2017neural, zhou2018deep, tang2018personalized}.

Note that the static recommendation module can be either general collaborative-based models \cite{he2017neural, koren2009matrix, he2020lightgcn} or sequential-based models \cite{kang2018self, gru4rec, tang2018personalized}. The primary distinction between collaborative-based and sequential-based models lies in the user modeling aspect. Collaborative-based models solely rely on user features, whereas sequential-based models incorporate both user features and the user's historical interactions to model the user representation.

\subsection{Dynamic Repeat-aware Module}
According to the observations from the empirical analysis and previous studies~\cite{benson2016modeling, anderson2014dynamics}, the repeat consumption time gap $\Delta t = t - t'$ is the most crucial factor, where $t$ is the current target time, and $t'$ is the history consumption time. To explicitly model the user repeat intention without any distribution assumption, we use neural ODE as the dynamic repeat-aware module to predict the user repeat intention. Our dynamic repeat-aware module consists of three components: (1) An encoder that encodes the static user and item representation into initial latent states for the input of the neural ODE. (2) Neural ODE for modeling the dynamic repeat-aware representation given the historical repeat time gaps and initial states. (3) A decoder that takes the repeat-aware representation as input and predicts the user's repeat intention.

\subsubsection{Encoder for Initial State} To capture a user-specific and item-specific repeat intention, we first transform the user representation $\boldsymbol{e}_u$ and item representation $\boldsymbol{e}_i$ from the static recommendation module into the initial latent representations by an $\textbf{Encoder}$: 
\begin{equation}
    \boldsymbol{h}_{u,i}^0 = \textbf{Encoder} (\boldsymbol{e}_u \| \boldsymbol{e}_i),
\end{equation}
where operator `$\|$' denotes the concatenation of two vectors. We implement \textbf{Encoder} with Multi-Layer Perceptron (MLP).

\subsubsection{Neural ODE for Repeat-aware Representation} 
Given the initial state, we use neural ODE to model the repeat-aware representation at the historical repeat time gaps:
\begin{equation}\label{eq:NODE_rep}
    \boldsymbol{h}_{u,i}^1, \ldots, \boldsymbol{h}_{u,i}^n = \textbf{ODESolve}\big(f_\text{ODE}, \boldsymbol{h}_{u,i}^0, (\Delta t_1, \ldots, \Delta t_n)  \big),
\end{equation}
where the $f_\text{ODE}$ is defined as another two-layer MLP to model the dynamics of the hidden state. Given the ODE function $f_\text{ODE}$, we can then leverage various ODE solvers to compute the hidden state at each time by integrating the ODE over the specified time interval. As mentioned in Section~\ref{sec:NODE}, a number of methods~\cite{runge1895numerische, dormand1980family} can be used to solve Eq.~\eqref{eq:NODE_rep}. In this paper, we use the Euler ODE solver~\cite{runge1895numerische}.

\subsubsection{Decoder for Prediction} After obtaining the repeat-aware representation for each user-item pair at the historical repeat time gaps, we predict the final user repeat intention through a $\textbf{Decoder}$:
\begin{equation}
    \hat{R}_{u,i}^\text{rep} =  \sum_{k=1}^n \textbf{Decoder}(\boldsymbol{h}_{u,i}^k), 
\end{equation}
where the $\textbf{Decoder}$ is also implemented with MLP.

\subsection{Model Learning}
After obtaining the final prediction score from Eq.~\eqref{eq: final_prediction}, we utilize pairwise ranking loss to optimize all the trainable parameters $\Theta$: 
\begin{equation}
\mathcal{L}_{\text{ReCODE}}(\Theta)=-\sum_{u \in \mathcal{U}} \sum_{n=1}^{N_u} \log \sigma\left(\hat{R}_{u,i_{n}}-\hat{R}_{u,i_{n}^-}\right),
\end{equation}
where we pair each ground-truth item $i_{n}$ with a negative item $i_{n}^-$ that is randomly sampled from items that users have not consumed.

\subsection{Discussion}


The dynamic repeat-aware module in our approach draws inspiration from neural ODE and does not rely on any specific distribution assumption. Indeed, note that the distribution assumptions made in existing studies \cite{dorogovtsev2000evolution, anderson2014dynamics, reiter2021predicting, wang2019modeling}, such as the Exponential distribution or Gaussian distribution, can be viewed as specific ordinary differential equations. For instance, the corresponding ODEs to Exponential distribution and Gaussian distribution are given by:
\begin{equation*}
     \frac{d\boldsymbol{h}(t)}{dt} = -\lambda \boldsymbol{h}(t) \quad~\text{and} \quad~ \frac{d\boldsymbol{h}(t)}{dt} = -\frac{(t-\mu)}{\sigma^2}\boldsymbol{h}(t),
\end{equation*}
respectively. $\lambda$ represents the rate parameter, and $\mu$ and $\sigma$ represent the mean and standard deviation, respectively.
By recognizing these distributions as ODE, we can see that our approach offers a more flexible and general framework for modeling repeat consumption dynamics, as it does not rely on specific distribution assumptions and can capture a wider range of temporal patterns.

\begin{table}[t]
\caption{Statistics of the experimental datasets.}
\label{tab:stat}
\resizebox{0.85\columnwidth}{!}{
\begin{tabular}{ccccc}
\toprule
\multicolumn{1}{c}{Dataset} & \multicolumn{1}{c}{\# user} & \multicolumn{1}{c}{\# item} & \multicolumn{1}{c}{\# interaction} & \multicolumn{1}{c}{repeat ratio} \\
\midrule
MMTD             & $10,150$     & $12,815$     & $256,315 $  & $21.40\%$      \\
Nowplaying-RS    & $44,971$     & $33,595$     & $2,240,871$  & $34.72\%$      \\
\bottomrule
\end{tabular}}
\end{table}

\section{Experiments}
In this section, we conduct experiments on two real-world datasets. The code is provided in \url{https://github.com/quchangle1/ReCODE}.

\subsection{Experimental Settings}

\subsubsection{Datasets} To evaluate the effectiveness of our proposed ReCODE, we conduct experiments on two real-world recommendation datasets: \textbf{MMTD\footnote{\href{http://www.cp.jku.at/datasets/MMTD/}{http://www.cp.jku.at/datasets/MMTD/}}}~\cite{hauger2013million} and \textbf{Nowplaying-RS\footnote{\href{https://zenodo.org/record/3247476\#.Yhnb7ehBybh}{https://zenodo.org/record/3247476\#.Yhnb7ehBybh}}}~\cite{poddar2018nowplaying}. 
These two datasets are from different platforms for music recommendation, where the repeated listening behavior of users is significant.
The statistics of both datasets are summarized in \autoref{tab:stat}.

\subsubsection{Evaluation Metrics} 

To evaluate the performance, we adopt the widely used ranking metrics Recall@$K$ and NDCG@$K$ and report the metrics for $K\in \{50, 100\}$. Following the common practice~\cite{kang2018self, sun2019bert4rec, he2017neural}, we apply the leave-one-out splitting strategy for evaluation. To better verify the effectiveness of our method, we calculate all metrics according to the rank of the ground-truth item among all items as candidates.

\subsubsection{Implementation Details} 
ReCODE is implemented based on the ReChorus~\cite{wang2020make} benchmark. We use the Euler ODE solver from the \textit{torchdiffeq}\footnote{\url{https://github.com/rtqichen/torchdiffeq}} package. 
To ensure a fair comparison, we set the embedding size of all methods to $32$ and use a batch size of $512$. The Adam~\cite{kingma2014adam} optimizer is employed to train all models, and we carefully tune the learning rate among $\{1e\text{--}3, 5e\text{--}4,1e\text{--}4,5e\text{--}5,1e\text{--}5\}$, as well as the weight decay among $\{1e\text{--}5, 1e\text{--}6, 1e\text{--}7\}$. For sequential-based methods, we set the maximum length of historical interactions to $20$. We run all models five times with different random seeds and report the average results.

\begin{table}
\centering
\caption{Performance $(\%)$ of ReCODE and baselines on two datasets, where R and N represent Recall and NDCG, respectively. `$**$' indicates the improvements over best baselines are statistically significant ($t$-test with $p$-value $< 0.01$).}
\vspace{-0.5em}
\label{tab: main_res}
\resizebox{1.\columnwidth}{!}{
\begin{tabular}{c|cccc|cccc}
\hline
\hline
Dataset    & \multicolumn{4}{|c}{MMTD}                                & \multicolumn{4}{|c}{Nowplaying-RS} \\
\hline
Model      & R@50    & N@50   & R@100   & N@100   & R@50       & N@50    & R@100   & N@100   \\
\hline
GRU4Rec    & $13.19$ & $4.05$ & $20.04$ & $5.15 $ & $19.74$  & $8.41 $  & $26.51$ & $9.51 $\\
Caser      & $13.37$ & $4.16$ & $20.17$ & $5.27 $ & $17.52$  & $7.24 $  & $24.16$ & $8.31 $\\
NARM       & $13.95$ & $4.24$ & $20.89$ & $5.36 $ & $20.24$  & $8.69 $  & $26.93$ & $9.77 $\\
SASRec     & $18.04$ & $6.90$ & $25.00$ & $8.03 $ & $22.74$  & $10.23$  & $29.62$ & $11.35$ \\
ContraRec  & $18.25$ & $6.11$ & $25.92$ & $7.35 $ & $20.38$  & $8.73 $  & $27.12$ & $9.82$ \\
\hline
ReCODE     & $\textbf{25.20}^{**}$ & $\textbf{9.58}^{**}$ & $\textbf{31.80}^{**}$ & $\textbf{10.64}^{**}$ & $\textbf{29.28}^{**}$  & $\textbf{12.84}^{**}$  & $\textbf{35.31}^{**}$ & $\textbf{13.82}^{**}$ \\
\hline
\hline
\end{tabular}}
\end{table}

\begin{table}
\centering
\caption{Performance $(\%)$ of ReCODE and SLRC when equipped with different base models on two datasets.
}
\vspace{-0.5em}
\label{tab:res_agnostic}
\resizebox{1\columnwidth}{!}{
\begin{tabular}{l|cccc|cccc}
\hline
\hline
Dataset    & \multicolumn{4}{|c}{MMTD}                                & \multicolumn{4}{|c}{Nowplaying-RS} \\
\hline
Model             & R@50    & N@50    & R@100   & N@100   & R@50       & N@50    & R@100   & N@100   \\
\hline
MF                & $18.68$ & $5.93$  & $26.61$ & $7.22 $ & $18.96$  & $7.92 $ & $26.08$ & $9.07 $ \\
+ SLRC            & $21.83$ & $9.11$  & $29.42$ & $10.34$ & $24.62$  & $9.46 $ & $30.66$ & $12.80$ \\
+ ReCODE          & $\textbf{25.20}^{**}$ & $\textbf{9.58}^{**}$  & $\textbf{31.80}^{**}$ & $\textbf{10.64}^{**}$ & $\textbf{29.28}^{**}$  & $\textbf{12.84}^{**}$ & $\textbf{35.31}^{**}$ & $\textbf{13.82}^{**}$ \\
\hline
NCF               & $17.33$ & $5.28 $ & $25.57$ & $6.62 $ & $19.56$  & $8.08 $ & $26.51$ & $9.20 $ \\
+ SLRC            & $20.11$ & $7.77 $ & $27.41$ & $8.96 $ & $23.90$  & $11.37$ & $30.07$ & $12.37$ \\
+ ReCODE          & $\textbf{24.41}^{**}$ & $\textbf{9.43}^{**} $ & $\textbf{30.63}^{**}$ & $\textbf{10.43}^{**}$ & $\textbf{29.56}^{**}$  & $\textbf{12.92}^{**}$ & $\textbf{35.42}^{**}$ & $\textbf{13.87}^{**}$ \\
\hline
GRU4Rec           & $13.19$ & $4.05 $ & $20.04$ & $5.15 $ & $19.74$  & $8.41 $ & $26.51$ & $9.51 $\\
+ SLRC            & $16.80$ & $7.53 $ & $22.42$ & $8.43 $ & $23.88$  & $11.47$ & $29.64$ & $12.41$ \\
+ ReCODE          & $\textbf{23.07}^{**}$ & $\textbf{9.26}^{**} $ & $\textbf{28.44}^{**}$ & $\textbf{10.12}^{**}$ & $\textbf{28.78}^{**}$  & $\textbf{12.94}^{**}$ & $\textbf{34.51}^{**}$ & $\textbf{13.87}^{**}$ \\
\hline
SASRec            & $18.04$ & $6.90 $ & $25.00$ & $8.03 $ & $22.74$  & $10.23$ & $29.62$ & $11.35$ \\
+ SLRC            & $19.06$ & $8.10 $ & $25.70$ & $9.18 $ & $24.53$  & $11.78$ & $30.68$ & $12.77$ \\
+ ReCODE          & $\textbf{24.86}^{**}$ & $\textbf{9.94}^{**} $ & $\textbf{30.76}^{**}$ & $\textbf{10.89}^{**}$ & $\textbf{30.20}^{**}$  & $\textbf{13.51}^{**}$ & $\textbf{36.38}^{**}$ & $\textbf{14.51}^{**}$ \\
\hline
\hline
\end{tabular}}
\end{table}

\subsection{Results and Analysis}
\subsubsection{Overall Performance}
We first implement our proposed ReCODE with MF~\cite{koren2009matrix} as the static recommendation module, and compare it with other competitive sequential recommendation models, including GRU4Rec~\cite{gru4rec}, Caser~\cite{tang2018personalized}, NARM~\cite{li2017neural}, SASRec~\cite{kang2018self} and ContraRec~\cite{wang2023sequential}. \autoref{tab: main_res} reports the averaged results of ReCODE and the baselines on MMTD and Nowplaying-RS datasets after five different runs. Remarkably, despite the simplicity of the static recommendation module in ReCODE, which employs MF, it consistently outperforms other sequential baselines, even those incorporating attention or transformer architectures such as NARM, SASRec, and ContraRec.  These findings underscore the critical role of modeling repeated consumption in real-world recommender systems, particularly in scenarios with high repetition rates.

\subsubsection{Model-Agnostic}
Given that proposed ReCODE is model-agnostic, we extend its applicability by integrating it with several popular recommendation models (called base models) to verify the effectiveness: (i) collaborative-based models: MF~\cite{koren2009matrix} and NCF~\cite{he2017neural}; (ii) sequential-based models: GRU4Rec~\cite{gru4rec} and SASRec~\cite{kang2018self}. Additionally, we compare ReCODE with another model-agnostic repeat-aware model SLRC~\cite{wang2019modeling}, which also can be applied to different base models. As shown in \autoref{tab:res_agnostic}, all base models equipped with ReCODE consistently improve their performance and outperform the baselines across all metrics on both datasets. 
These results unequivocally validate ReCODE's effectiveness in enhancing a wide range of base models, including those based on collaborative filtering and sequential recommendation techniques.

\subsubsection{Performance w.r.t. Different Repeat Ratios}
\begin{figure}[t]
  \centering
  \vspace{-0.4cm}
  \subfigcapskip=-7pt
  \subfigure[MMTD]
      {
    \includegraphics[width=0.47\textwidth]{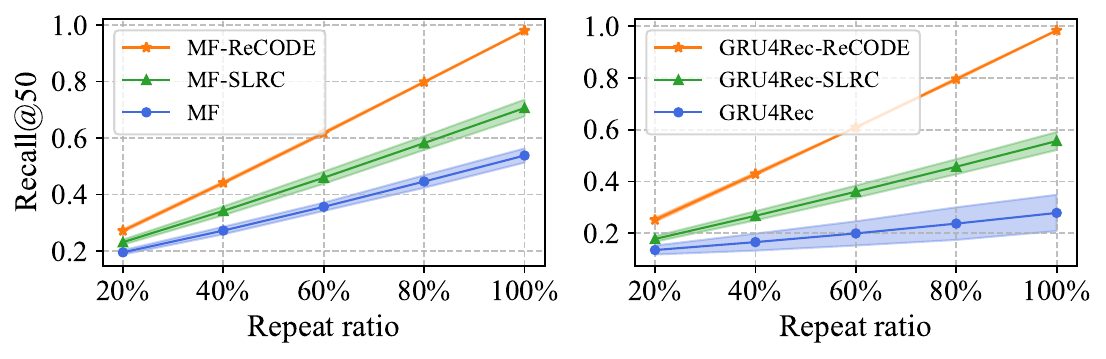}
    \label{fig:mmtd_ana}
  }
  \subfigcapskip=-7pt
  \subfigure[Nowplaying-RS]{
    \includegraphics[width=0.47\textwidth]{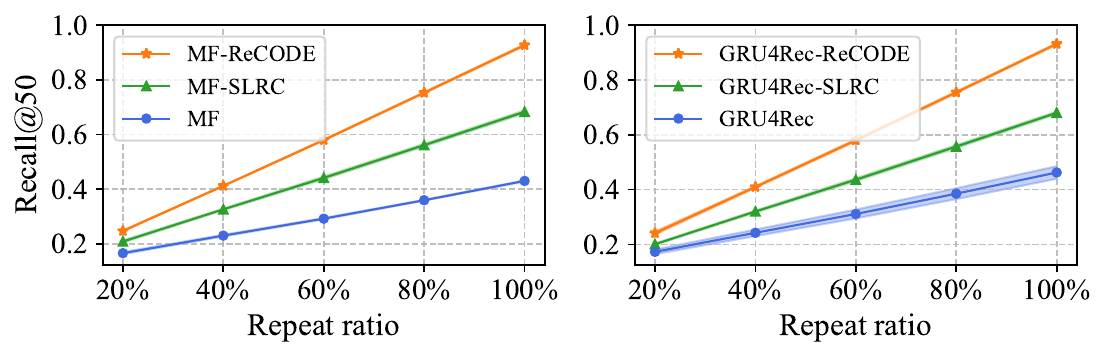}
    \label{fig:nowplaying_ana}
  }
  \caption{Performance comparisons (Recall@50) w.r.t. different repeat ratios on MMTD and Nowplaying-RS datasets. The shaded area represents the 95\% confidence intervals of $t$-distribution obtained from the five experimental runs.}
    \label{fig:exp_ana}
\end{figure}

To further investigate the influence of datasets with different repeat ratios, we conduct experiments using MF and GRU4Rec as base models.
\autoref{fig:exp_ana} shows the results on both MMTD and Nowplaying-RS datasets. It is evident that ReCODE consistently achieves superior performance under all circumstances and exhibits a higher improvement as the repeat ratios increase. This result further highlights the ability of ReCODE to effectively capture the dynamic patterns of repeat consumption, resulting in enhanced recommendation performance compared to the baselines.

In summary, the above results from \autoref{tab: main_res}, \autoref{tab:res_agnostic},  and \autoref{fig:exp_ana} collectively confirm the effectiveness of ReCODE in enhancing the recommendation performance of different base models. We attribute this to the superiority and stability of ReCODE in modeling repeat consumption, which makes it a promising approach for improving recommender systems in real-world scenarios, particularly in cases with high repeat ratios.

\section{Conclusion}
This paper proposes a novel neural ODE-based approach to modeling the repeat consumption in recommendation, called ReCODE. By leveraging neural ODE, ReCODE effectively captures the dynamic patterns of repeat time gaps without making any distribution assumptions. Additionally, ReCODE is model-agnostic, making it applicable to various base models, including both collaborative-based and sequential-based models.  Experimental results on two real-world datasets and four base models demonstrate the effectiveness of ReCODE in improving recommendation performance.

\begin{acks}
This work was funded by the National Key R\&D Program of China (2023YFA1008704), the National Natural Science Foundation of China (No. 62377044, No.62376275), Beijing Key Laboratory of Big Data Management and Analysis Methods, Major Innovation \& Planning Interdisciplinary Platform for the  ``Double-First Class” Initiative, PCC@RUC, funds for building world-class universities (disciplines) of Renmin University of China. 

\end{acks}


\bibliographystyle{ACM-Reference-Format}
\balance
\bibliography{mybib}


\end{document}